\documentclass[aps,prd,twocolumn,groupedaddress]{revtex4}
\usepackage{graphicx}
\newcommand{\be}{\begin{equation}}
\newcommand{\ee}{\end{equation}}
\newcommand{\ba}{\begin{eqnarray}}
\newcommand{\ea}{\end{eqnarray}}
\newcommand{\Eq}[1]{Eq.~(\ref{#1})}

\newcommand{\Fig}[1]{Fig.~{\ref{#1}}}

\newcommand{\Table}[1]{Table~{\ref{#1}}}
\begin{document}

\title{Covariant QCD Modeling of Light Meson Physics} 
\thanks{Presented at International School on Nuclear Physics, Erice, September
2002; to appear in Prog. Part. Nucl. Phys.}
\author{Peter~C.~Tandy}
\email[]{tandy@cnr2.kent.edu}
\affiliation{Center for Nuclear Reseach, Department of Physics,
             Kent State University, Kent, Ohio 44242 U.S.A.}
\date{January 7, 2003}
\begin{abstract}
We summarize recent progress in soft QCD modeling based on the 
set of Dyson--Schwinger equations truncated to ladder-rainbow level.   
This covariant approach to hadron physics accommodates quark confinement 
and implements the QCD one-loop renormalization group behavior.  We 
compare the dressed quark propagator, pseudoscalar and vector 
meson masses as a function of quark mass, and the \mbox{$\rho \to \pi \pi$} 
coupling to  recent lattice-QCD data.  The error in the 
Gell-Mann--Oakes--Renner relation with increasing quark mass is quantified
by comparison to the exact pseudoscalar mass relation as evaluated 
within the ladder-rainbow  Dyson-Schwinger model.
\end{abstract}


%

\maketitle

\section{Introduction}

The study of light-quark pseudoscalar and vector mesons is an
important tool for understanding how QCD works in the non-perturbative
regime.  The pseudoscalars are important because they are the lightest
observed hadrons and are the Goldstone bosons associated with
dynamical chiral symmetry breaking.  The ground state vector mesons
are important because, as the lowest spin excitations of the pseudoscalars, 
they relate closely to hadronic $\bar{q}q$ modes that are electromagnetically 
excited.

We use a Poincar\'e covariant model
defined within the framework of the
Dyson--Schwinger equations [DSEs] of QCD; these form an excellent tool to
study nonperturbative aspects of hadron properties~\cite{Roberts:2000aa}.  
It is straightforward to implement the correct one-loop
renormalization group behavior of QCD~\cite{Maris:1997tm}, and obtain
agreement with perturbation theory in the perturbative region.
Provided that the relevant Ward--Takahashi identities  are
preserved in the truncation of the DSEs, the corresponding currents
are conserved.  Axial current conservation induces the Goldstone
nature of the pions and kaons~\cite{Maris:1998hd}; electromagnetic
current conservation produces the correct electric charge of the
mesons without
fine-tuning.  These properties are implemented here within the rainbow
truncation of the DSE for the dressed quark propagators together with
the ladder approximation for the Bethe--Salpeter equation [BSE] for
meson bound states.  

The model~\cite{Maris:1999nt} we use has two infrared parameters 
which specify the momentum 
distribution and strength of the ladder-rainbow kernel at a low scale 
necessary to generate an empirically acceptable amount of dynamical chiral 
symmetry breaking~\cite{Atkinson:1988mw,Roberts:1990mj} as measured by 
the chiral condensate.   As a corollary, the strong dressing of the quark 
propagator shifts the mass pole significantly away from the real timelike
$p^2$ axis.   The produced bound state mesons do not have a $\bar q q$ 
decay width and, in this sense, the present model implements quark
confinement.   The absence of a real mass pole for dressed quark and 
gluon propagators has been studied and found to be a sufficient, but not
necessary, condition for 
confinement~\cite{Roberts:2000aa,Burden:1992gd,Krein:1992sf,Maris:1995ns}.
The model provides an efficient description of the masses and decay
constants of the light-quark 
pseudoscalar and vector mesons~\cite{Maris:1997tm,Maris:1999nt},  the elastic 
charge form factors $F_\pi(Q^2)$ and $F_K(Q^2)$~\cite{Maris:2000sk} and 
the electroweak transition form
factors of the pseudoscalars and vectors~\cite{Maris:2002mz,Ji:2001pj}.


%
\section{Dyson-Schwinger Equations}

The dressed quark propagator $S(p)$ is the solution to
\ba
S(p)^{-1}\;&&= \;Z_2 \, i\,/\!\!\!p + Z_4 \, m(\mu) \nonumber \\ 
     &&+Z_1 \int^\Lambda_q \! g^2D_{\mu\nu}(p-q) \, 
        \frac{\lambda^i}{2}\gamma_\mu \, S(q) \, \Gamma^i_\nu(q,p)~,
\label{quarkdse}
\ea
where $D_{\mu\nu}(k)$ is the renormalized dressed-gluon propagator,
$\Gamma^i_\nu(q,p)$ is the renormalized dressed quark-gluon vertex.
The notation \mbox{$\int^\Lambda_q = $} 
\mbox{$\int^\Lambda d^4q/(2\pi)^4$} denotes a translationally invariant
regularization of the integral with mass-scale $\Lambda$.
The solution  is renormalized according to
$S(p)^{-1}=i\gamma\cdot p+m(\mu)$ at a sufficiently large spacelike
$\mu^2$, with $m(\mu)$ the renormalized quark mass at the scale $\mu$.
The renormalization constants $Z_2$ and $Z_4$ depend on the
renormalization mass-scale $\mu$ and on the regularization mass-scale
$\Lambda$.  The limit \mbox{$\Lambda \to \infty$} is to be taken at
the end of all calculations.

The BSE for a $a\bar b$ meson is
\be
\Gamma^{a\bar{b}}(p_+,p_-) = \int^\Lambda_q \! K(p,q;P)S^a(q_+)
                                   \Gamma^{a\bar{b}}(q_+,q_-)S^b(q_-),
\label{bse}
\ee
where $K$ is the renormalized $q\bar{q}$ scattering kernel that is
irreducible with respect to a pair of $q\bar{q}$ lines. The quark
momenta are $q_\pm$; the meson momentum is \mbox{$P= q_+ - q_-$} and
satisfies $P^2 = -m^2$.  The relative momentum $q$ is introduced
by \mbox{$q_+ =$} \mbox{$q+\eta P$} and \mbox{$q_- =$} \mbox{$q -$}
\mbox{$(1-\eta) P$} where $\eta$ is the momentum partitioning
parameter.  Physical observables should not depend on $\eta$ and
this provides a convenient check on numerical methods.
\begin{figure}[htb]
\includegraphics[width=8.5cm]{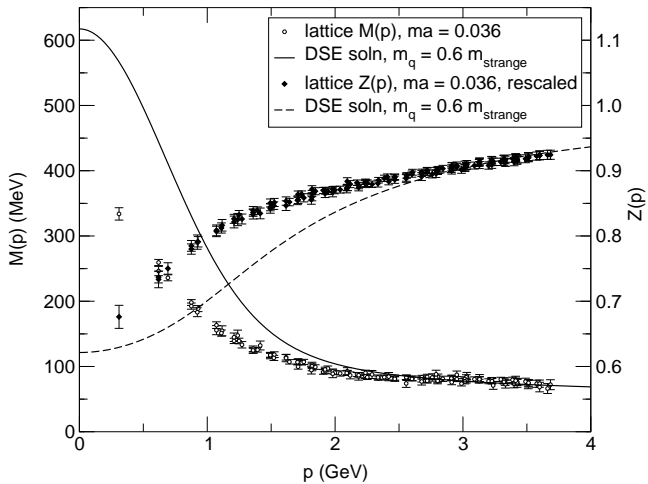}
\caption{DSE solution~\protect\cite{Maris:1999nt,Jarecke:2002xd} 
for quark propagator amplitudes compared to recent lattice
data~\protect\cite{Bowman:2002bm,Bowman_privCom02}.  
\label{fig:MplusZ} }
\end{figure}
We employ the model that has been developed recently for an efficient
description of the masses and decay constants of the light
pseudoscalar and vector mesons~\cite{Maris:1997tm,Maris:1999nt}.  This
consists of the rainbow truncation of the DSE for the quark propagator
and the ladder truncation of the BSE for the pion and kaon amplitudes.
The required effective $\bar q q$ interaction is constrained by
perturbative QCD in the ultraviolet and has a phenomenological
infrared behavior.  In particular, the rainbow truncation of the quark
DSE, Eq.~(\ref{quarkdse}), and the ladder truncation of the BSE,
Eq.~(\ref{bse}), are
\be
\label{ourDSEansatz}
Z_1 g^2 D_{\mu \nu}(k) \Gamma^i_\nu(q,p) \rightarrow
 4\pi \alpha_{\rm eff}(k^2) D_{\mu\nu}^{\rm free}(k)\, \gamma_\nu
                       \textstyle\frac{\lambda^i}{2}\;,
\ee
and
\be
K(p,q;P) \to
        - 4\pi \alpha_{\rm eff}(k^2) D_{\mu\nu}^{\rm free}(k)\,
        \textstyle{\frac{\lambda^i}{2}}\gamma_\mu \otimes
        \textstyle{\frac{\lambda^i}{2}}\gamma_\nu \,,
\ee
where $D_{\mu\nu}^{\rm free}(k=p-q)$ is the free gluon propagator in
Landau gauge.  These two truncations are consistent in the sense
that the combination produces vector and axial-vector vertices
satisfying the respective WTIs.  In the axial case, this ensures that in
the chiral limit the ground state pseudoscalar mesons are the massless
Goldstone bosons associated with chiral symmetry
breaking~\cite{Maris:1997tm,Maris:1998hd}.  In the vector case, this ensures
electromagnetic current conservation.
The ``effective coupling'' $\alpha_{\rm eff}(k^2)$ defines the model.  
The ultraviolet behavior is
chosen to be that of the QCD running coupling $\alpha(k^2)$; the
ladder-rainbow truncation then generates the correct perturbative QCD
structure of the DSE-BSE system of equations.  The phenomenological
infrared form of $\alpha_{\rm eff}(k^2)$ is chosen so that the DSE kernel
contains sufficient infrared enhancement to produce an empirically
acceptable amount of dynamical chiral symmetry breaking as represented
by the chiral condensate~\cite{Hawes:1998cw}.

We employ the Ansatz found to be successful in earlier 
work~\cite{Maris:1997tm,Maris:1999nt}
\be
\label{gvk2}
\frac{{\cal G}(k^2)}{k^2} =
        \frac{4\pi^2\, D \,k^2}{\omega^6} \, {\rm e}^{-k^2/\omega^2}
+ \frac{ 4\pi^2\, \gamma_m \; {\cal F}(k^2)}
        {\frac{1}{2} \ln\left[\tau +
        \left(1 + k^2/\Lambda_{\rm QCD}^2\right)^2\right]} \,,
\ee
{\sloppy 
with \mbox{$\gamma_m=\frac{12}{33-2N_f}$} and
\mbox{${\cal F}(s)=$} \mbox{$(1 - \exp(\frac{-s}{4 m_t^2}))/s$}.
The first term implements the strong infrared enhancement in the region
\mbox{$0 < k^2 < 1\,{\rm GeV}^2$} required for sufficient dynamical
chiral symmetry breaking.  The second term serves to preserve the
one-loop renormalization group behavior of QCD.  We use
\mbox{$m_t=0.5\,{\rm GeV}$}, \mbox{$\tau={\rm e}^2-1$}, \mbox{$N_f=4$},
and we take \mbox{$\Lambda_{\rm QCD} = 0.234\,{\rm GeV}$}.  The
renormalization scale is chosen to be \mbox{$\mu=19\,{\rm GeV}$} which
is well into the domain where one-loop perturbative behavior is
appropriate~\cite{Maris:1997tm,Maris:1999nt}.  The remaining parameters,
\mbox{$\omega =$} \mbox{$0.4\,{\rm GeV}$} and \mbox{$D=0.93$} 
\mbox{${\rm GeV}^2$} along with the quark masses, are fitted to give 
a good description of $\langle\bar q q\rangle$, $m_{\pi/K}$ and
$f_{\pi}$ as shown in \Table{model_fit}.
\begin{table}
\caption{\label{model_fit}
The pseudoscalar observables that define the present ladder-rainbow DSE-BSE
model, adapted from Refs.~\protect\cite{Maris:1997tm,Maris:1999nt}.  
}
\begin{ruledtabular}
\begin{tabular}{l|cc}
  & \multicolumn{1}{r}{experiment}
  & \multicolumn{1}{r}{calculated}  \\
  & \multicolumn{1}{r}{(estimates)}
  & \multicolumn{1}{r}{($^\dagger$ fitted)} \\ \hline
$m^{u=d}_{\mu=1 {\rm GeV}}$ &
   \multicolumn{1}{r}{ 5 - 10 MeV}  & \multicolumn{1}{r}{ 5.5 MeV}     \\
$m^{s}_{\mu=1 {\rm GeV}}$ &
   \multicolumn{1}{r}{ 100 - 300 MeV} &\multicolumn{1}{r}{ 125 MeV} \\ \hline
- $\langle \bar q q \rangle^0_{\mu}$
                & (0.236 GeV)$^3$ & (0.241$^\dagger$)$^3$ \\
$m_\pi$         &  0.1385 GeV &   0.138$^\dagger$ \\
$f_\pi$         &  0.131 GeV &   0.131$^\dagger$ \\
$m_K$           &  0.496 GeV  &   0.497$^\dagger$ \\
$f_K$           &  0.160 GeV  &   0.155     \\ 
\end{tabular}
\end{ruledtabular}
\end{table}
The subsequent values for $f_K$ and the masses and decay
constants of the vector mesons $\rho, \phi, K^\star$ are found to be
within 10\% of the experimental data~\cite{Maris:1999nt}.  
\begin{figure}[htb]
\hspace*{-2mm}
\includegraphics[width=7.5cm,angle=-90]{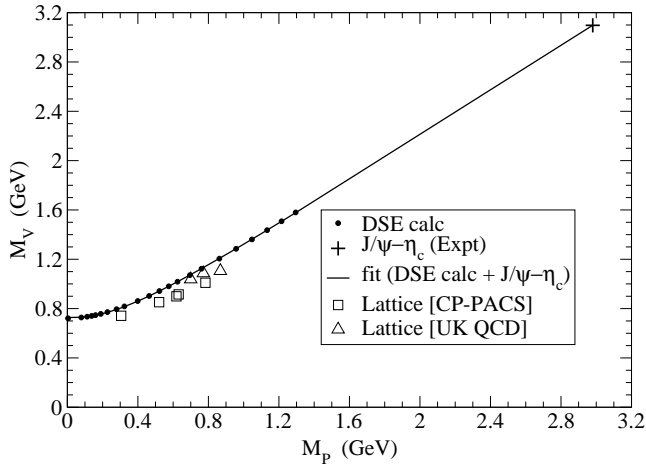}
\caption{DSE calculation of equal flavor vector meson mass 
variation with pseudoscalar meson mass as $m_q$ is varied compared
to lattice data from CP-PACS~\protect\cite{Aoki:1999ff} and 
UK-QCD~\protect\cite{Allton:1998gi}.
\label{fig:MVvsMP} }
\end{figure}

In \Fig{fig:MplusZ}  we compare the
DSE model propagator amplitudes defined by
\mbox{$S(p) =$} \mbox{$Z(p^2)[ i /\!\!\! p +$} \mbox{$M(p^2)]^{-1}$} with the
most recent results in lattice QCD using staggered fermions in Landau
gauge~\cite{Bowman:2002bm,Bowman_privCom02}.  These simulations were
done with the Asqtad improved staggered quark action, which has
lattice errors of order ${\cal O}(a^4)$ and ${\cal O}(a^2\,g^2)$.  
\Fig{fig:MplusZ} shows both $M(p)$ and
$Z(p)$ obtained with a bare lattice mass of $ma = 0.036$ in lattice
units, which corresponds to a bare mass of $57$~MeV in physical units.
The DSE calculations use a current mass value of $75$~MeV at $\mu$ =
$1$~GeV to match the lattice mass function around $3$~GeV; this
current mass is about $0.6 \, m_s$.  There is agreement in the
qualitative infrared structure of $M(p)$ and $Z(p)$.   Since the lattice
simulation produces the regulated but un-renormalized propagator, the
scale of $Z(p)$ is arbitrary and we have
rescaled the lattice $Z(p)$ to match the DSE solution
at $3$~GeV.  For $Z(p)$ the ladder-rainbow DSE model  saturates much 
slower than does the lattice; this may signal
a deficiency of the bare gluon-quark vertex.   A recent study of the
coupled ghost-gluon-quark DSEs has found that the quark-gluon vertex 
dressing can produce a change of this character in the infrared 
structure of the quark amplitudes~\cite{Fischer_PhD02,Alkofer_privcom02}.

In \Fig{fig:MVvsMP} we compare the rainbow-ladder DSE model with 
unquenched lattice data for
the variation of vector and pseudoscalar meson masses with quark 
current mass in the case of equal flavor quarks.  The DSE calculation,
shown by the discrete circles, is limited to the mass range where it is 
reliable.   The solid line is a fit to those results plus the 
experimental $J/\psi-\eta_c$ point.  The curvature
at low mass is consistent with \mbox{$M_V \propto m_q$} and
\mbox{$M_P \propto \sqrt{m_q}$}.  This comparison is consistent with
the known properties of the DSE results:
$M_V$ is 5\% too low for the $\rho$ and 5\% too high for 
the $\phi$, while the pseudoscalar masses in the u-quark and s-quark regions
are fit to experiment.  (If the evident fit is continued to the $\Upsilon$ 
vector mass, the predicted $\eta_b$ mass would be 10.0~GeV.)

%
\begin{figure}[htb]
\hspace*{1mm}
\includegraphics[width=7.0cm,angle=-90]{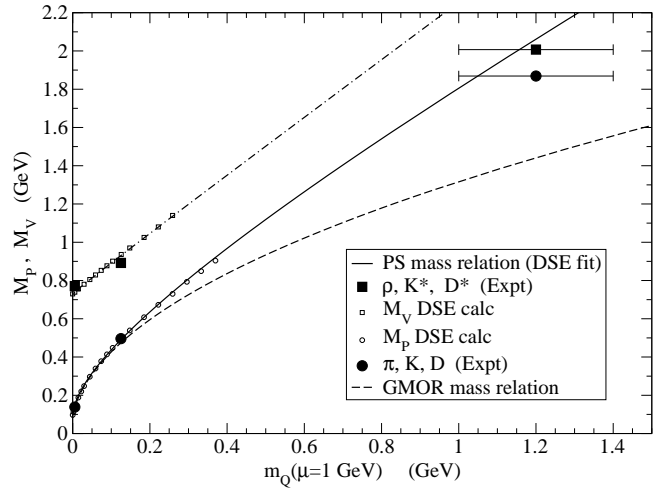}
\caption{$M_V(m_Q)$ and $M_P(m_Q)$ for unequal 
flavor $u-Q$ mesons.  The DSE model-exact 
mass relation reveals the size of the correction to the GMOR relation.  
\label{fig:MvsmQ} }
\end{figure}
\section{Pseudoscalar Meson Mass Relation}

As the current quark mass is raised from zero, the explicit breaking of 
chiral symmetry adds mass to the Goldstone boson modes.   The way in
which the pseudoscalar meson mass grows with quark mass is described,
at low mass, by the GMOR relation.   This is
\be
M^2_P\big(m_1(\mu), m_2(\mu)\big) = [m_1(\mu) + m_2(\mu)]\; 
                      \frac{|\langle \bar q q \rangle^0_\mu|}
                           {(f_P^0)^2} + {\cal O}(m^2)~~,
\ee
in the general case where the two quark flavors are different.  Here
\mbox{$\langle \bar q q \rangle^0_\mu =$} 
\mbox{$ -Z_4 N_c{\rm tr}\int^\Lambda_q S_0(q)$} is the chiral
condensate at scale $\mu$, the current masses are determined
at the same scale, and $f_P^0$ is the chiral limit electroweak decay 
constant (in the \mbox{$f_\pi=$} 92.4 MeV convention).    For $u/d$ 
quarks the GMOR relation
is satisfied to high accuracy (within 0.2\% in the present DSE model).  For 
current masses of the order of
\mbox{$m_s \sim 120$}~MeV and above, the question of the size of the error
in the GMOR relation is not settled.  
\begin{figure}[htb]
\hspace*{2mm}
\includegraphics[width=6.8cm,angle=-90]{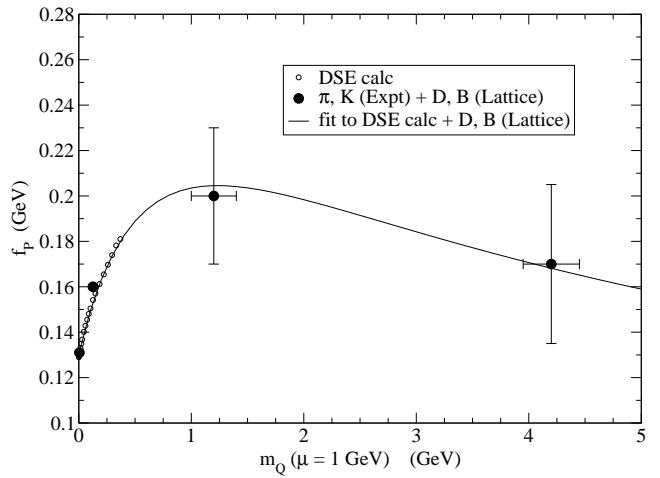}
\caption{Current quark mass dependence of $f_P$, the electroweak
decay constant for $u-Q$ mesons. [$f_\pi = 131$~MeV convention.]
\label{fig:fPvsmQ} }
\end{figure}
An exact mass relation for pseudoscalar mesons in QCD, applicable
for all values of the quark masses, has been 
established~\cite{Maris:1998hd}.  With allowance 
for different quark flavors, it takes the form
\be
M^2_P \; f_P(m_1, m_2) = (m_1 + m_2)\; R_P(m_1, m_2)~~,
\label{exactM}
\ee
where $R_P$ is the projection of the meson wave function onto $\gamma_5$
at the origin of $\bar q q$ separation and is given by~\cite{Maris:1998hd}
\be
   R_P = 
    -i\, Z_4 \, N_c \!\int^\Lambda_q \!\!{\rm tr}\big[ \gamma_5 
                S_{f_1}(q_+) \Gamma_P(q;P) S_{f_2}(q_-)\big]\, ,
\label{eq:Rp}
\ee
with all renormalized quantities taken at the same scale $\mu$.
The chiral limit of this quantity can be shown to be
\mbox{$R_P(m_1=m_2=0) = -\langle \bar q q \rangle^0_\mu/f_P^0$},
and thus the GMOR relation follows as a collorary of the exact
relation, \Eq{exactM}, at low mass.   The origin of this exact mass 
relation is the axial vector Ward-Takahashi relation
\ba
-i P_\mu \; \Gamma_{5\mu}(q;P) =&& S^{-1}_{f_1}(q_+)\, \gamma_5 +
\gamma_5 \,S^{-1}_{f_2}(q_-) \nonumber \\ 
&& - (m_1 + m_2)\; \Gamma_5(q;P)~~.
\label{AV_WTI}
\ea
The dressed vertices $\Gamma_{5\mu}(q;P)$ and $\Gamma_5(q;P)$
satisfy inhomogeneous integral equations that have the same kernel,
the irreducible 
$\bar q q$ scattering amplitude; the inhomogeneous terms are 
$Z_2\, \gamma_\mu$ and $Z_4\, \gamma_5$ respectively.   Thus both 
vertices have
poles corresponding to the pseudoscalar meson bound states.
[The axial vector poles in $\Gamma_{5\mu}$ have transverse residues and do 
not contribute.]  The exact mass relation, \Eq{exactM}, arises from 
the equality of
the pseudoscalar pole residues from both sides of \Eq{AV_WTI}.  The
residue of $\Gamma_5(q;P)$ is $-i\,R_P\, \Gamma_P(q;P)$, and the
residue of $\Gamma_{5\mu}(q;P)$ is $P_\mu\, f_P\, \Gamma_P(q;P)$, where
$\Gamma_P$ is the pseudoscalar bound state BS amplitude.  The expression
for $f_P P_\mu$ is the same as for $R_P$ in \Eq{eq:Rp} 
except that $\gamma_5$ is replaced by $\gamma_5 \gamma_\mu$ and 
$-i\, Z_4$ is replaced by $Z_2/\sqrt{2}$ (in the convention where the
physical $f_\pi$ is $92.4$~MeV).
\begin{figure}[htb]
\includegraphics[width=7.4cm,angle=-90]{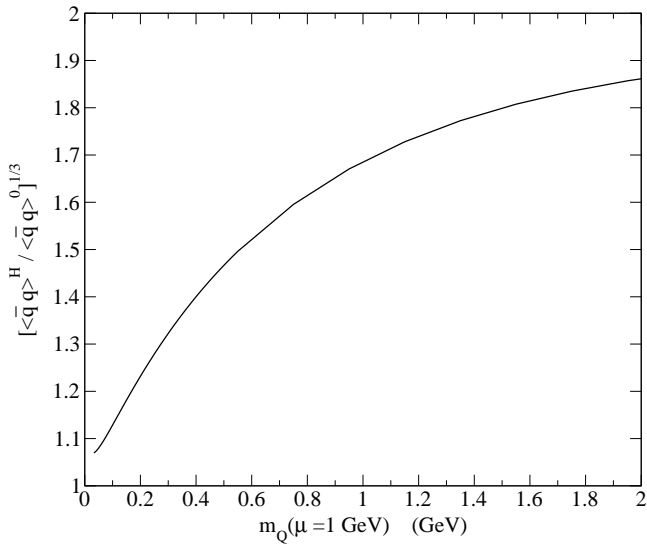}
\caption{A measure of the current quark mass dependence 
of the in-hadron condensate for $qQ$ pseudoscalars.  
\label{fig:cond_H} }
\end{figure}
In the chiral limit, the last term of \Eq{AV_WTI}
is not present and the right hand side is not singular.  A systematic 
expansion in powers of $P_\mu$ reveals~\cite{Maris:1998hd} that in 
leading order  the 
pole in the left hand side must move to  \mbox{$P^2=0$} to be 
cancelled and thus \mbox{$M_P=0$}.  The analysis also provides relationships
between the BS amplitude  and the quark propagator, e.g., 
\mbox{$f_P^0\; E_P^0(q;0) = B_0(q^2)$}, where  \mbox{$\Gamma_P(q;P)=$}
\mbox{$i\gamma_5 E_P + \cdots$}.  Since the quark mass function
is proportional to $B_0(q^2)$, this latter relation means that dynamical
chiral symmetry breaking is necessarily accompanied by a massless pseudoscalar
bound state.  This is the familar Goldstone's theorem; notice that the
composite, distributed nature of the pion amplitude requires a running 
quark mass function.

At small $m$, both $f_P$ and $R_P$ are constant leading
to the GMOR behavior \mbox{$M_P \sim \sqrt{m}$}.   The error in this 
has to increase with mass since the heavy quark
limiting behavior is~\cite{Ivanov:1998ms} \mbox{$f_P \sim 1/\sqrt{m}$} and
\mbox{$R_P \sim \sqrt{m}$} which leads to the linear behavior
\mbox{$M_P \sim m$}.    
\begin{figure}[htb]
\includegraphics[width=7.6cm,angle=-90]{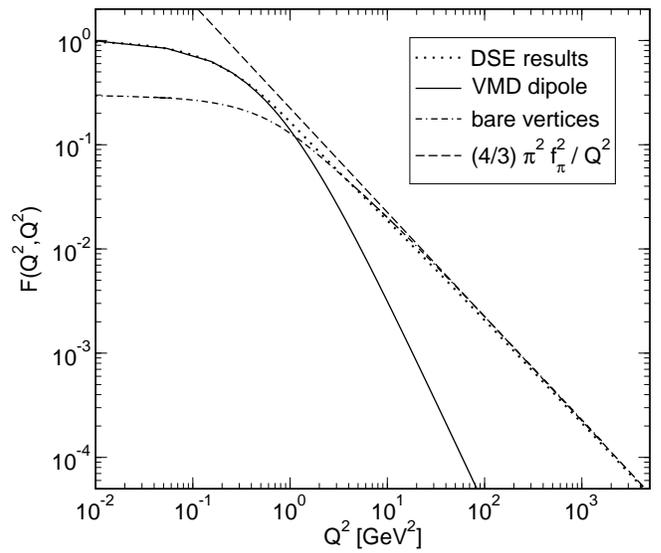}
\caption{Our DSE results for the symmetric $\gamma^* \pi \to \gamma^*$ form
factor, compared to the pQCD asymptotic $1/Q^2$ behavior. [Here
\mbox{$f_\pi=131~{\rm MeV}$}.]  The naive VMD model suggests a
dipole behavior which is correct only in the infrared.
\label{fig:piggsymlog} }
\end{figure}

The evolution of both vector and pseudoscalar  meson masses with 
increasing current quark mass has been studied within 
the DSE model in both the equal and unequal flavor cases.
An example is shown in \Fig{fig:MvsmQ} for $u-Q$ mesons as a 
function of $m_Q(\mu=1~{\rm GeV})$ up to the limit of accuracy of 
the calculations.   The fit shown is adapted from Ref.~\cite{Maris:2000zf} 
by evolving the $m_Q(\mu=19~{\rm GeV})$ used there to \mbox{$\mu=1~{\rm GeV}$}
for ease of comparison with conventionally quoted values.   Thus here we
have \mbox{$M_P = \alpha' +$} \mbox{$\beta' \sqrt{m_Q} +$}
\mbox{$\gamma' m_Q$};  with both masses in GeV, the parameters are
\mbox{$(\alpha', \beta', \gamma')=$} \mbox{$(0.083,0.842,0.880)$}.
With the DSE model-exact values of $f_P^0$ and 
$\langle \bar q q \rangle^0_\mu$,  the GMOR relation is explicitly
\mbox{$M^2_P=$} \mbox{$0.00955 +$} \mbox{$ 1.724\, m_Q(\mu=1~{\rm GeV})$}
and this is compared to the exact mass relation in \Fig{fig:MvsmQ}.   
For the $K$ meson the GMOR error is 4\%, at
$m_Q= 0.4$~GeV the error is 14\%, while at the D meson the error is 30\%.
In the D meson region, only $\sim 50$\% of the mass comes from the linear 
term; the heavy quark domain is at higher mass.  

The exact mass relation $M_P(m_Q)$, \Eq{exactM}, differs from the 
GMOR behavior due to 
the mass dependence of $f_P$ and $R_P$.  Instead of the latter quantity, 
one can define  \mbox{$\langle \bar q q \rangle^H_\mu = - f_P\; R_P$}
as an effective ``in-hadron'' condensate~\cite{Ivanov:1998ms} which 
allows the exact mass
relation to take the GMOR-like form \mbox{$M^2_P\, f_P^2 =$} 
\mbox{$ (m_1+m_2)_\mu\,|\langle \bar q q \rangle^H_\mu|$}.   From this 
relation we extract the quark mass dependence of 
$\langle \bar q q \rangle^H_\mu$
at $\mu=1$~GeV.  The low mass DSE results for $M_P$ and $f_P$ are fitted
to forms that respect the heavy quark limits, are consistent with $D$ and
$B$ meson masses, and are consistent with lattice 
results~\cite{Flynn:1998ca} for $f_P$
of the $D$ and $B$ mesons.   We adapt the $f_P(M_P)$ fit from 
Ref.~\cite{Ivanov:1998ms} to produce $f_P(m_Q)$ and to accommodate the 
DSE results at low mass.    The result is displayed in \Fig{fig:fPvsmQ}.
The fit (in the \mbox{$f_\pi =$} 131 MeV convention) is 
\mbox{$f_P^2 = N/D$}, where \mbox{$N= a + b\, m_Q$} and
\mbox{$D= 1.0 + c\, m_Q + d\, m_Q^2$}, with \mbox{$(a,b,c,d)= $}
\mbox{$(0.017, 0.068, 0.649, 0.391)$} when $m_Q$ is in GeV.  
The resulting estimate of the mass dependence
for $\langle \bar q q \rangle^H_\mu$ at $\mu=1$~GeV  is shown in 
\Fig{fig:cond_H} and indicates $\sim$ 15\% increase over the chiral limit 
value for an $m_Q$ relevant to $K$, while
for the $D$ meson the in-hadron condensate is about 70\% enhanced.

\section{Vector Meson Strong Decays}

Since this DSE model describes the elastic charge form factors of the
pseudoscalars very well~\cite{Maris:2000sk} in impulse
approximation,  the strong decays of the vector mesons should be 
well-described  without parameter adjustment.   In impulse 
approximation, the amplitude for the decay of $\rho$ with 4-momentum
$Q=p_1+p_2$ to $\pi\pi$ with 4-momenta $p_1, p_2$ is given 
by~\cite{Jarecke:2002xd}
\ba
2 P^T_\mu \; g_{\rho \to \pi \pi} =&& \sqrt{2} N_c {\rm tr} 
     \int^\Lambda_k \! S(q) \, \Gamma_P(q,q_+) \, S(q_+) \nonumber \\  
&& \times    \Gamma_\mu(q_+,q_-)\, S(q_-) \, \bar\Gamma_P(q_-,q) \;,
\label{triangle}
\ea
where no distinction is made between the u/d quarks,  
\mbox{$P=(p_1 - p_2)/2$}, \mbox{$q = $} \mbox{$k+P/2$} and 
\mbox{$q_\pm =$} \mbox{$ k-P/2 \pm$} \mbox{$ Q/2$}.   The appropriate 
generalizations
of the flavor structure appropriate to \mbox{$\phi \to K K$} and 
\mbox{$K^* \to K \pi$} are straightforward~\cite{Jarecke:2002xd}.  
In \Eq{triangle}, the  component of $P_\mu$ transverse to $Q_\mu$ is 
indicated on the left hand side to cover the case of unequal decay
products.    
\begin{table}
\caption{\label{tab:polefit} The coupling
constants $g_{v \to p p}$ calculated in the DSE model compared to results
from a fit to the timelike form factor 
pole~\protect\cite{Maris:2001rq,Maris:2001am} and 
lattice-QCD~\protect\cite{McNeile:2002fh}.
}
\begin{ruledtabular}
\begin{tabular}{c|cccc}
    $g_{v \to p p}$   & Expt    &   this work  & pole fit        
                        & lattice-QCD \\ \hline
$g_{\rho \to \pi\pi}$  & 6.02  &  5.14  & 5.2 & $6.08^{+2.04}_{-1.00}$  \\
$g_{\phi \to KK}$      & 4.64  &  4.25  & 4.3 &    -                   \\
$g_{K^{\star+}\to K^0 \pi^+}$  & 4.60 & 4.81 & 4.1 &    -          \\ 
%
\end{tabular}
\end{ruledtabular}
\end{table}

The results shown in Table~\ref{tab:polefit} are within  5-10\% 
of experiment with the error being larger if the
vector meson is lighter.  As an independent check, coupling
constants are also extracted from the timelike electroweak form factors
near the vector meson poles~\cite{Maris:2001rq,Maris:2001am}.  
The agreement is encouraging considering that with eight independent 
covariants for the vector BS amplitude and four each for the pseudoscalars, 
there are 128 distinct quark loop integrals for each physical decay.
Also shown in Table~\ref{tab:polefit} is a recent 
lattice-QCD result~\cite{McNeile:2002fh} for \mbox{$\rho \to \pi\pi$}
from the UKQCD Collaboration.  Although the lattice data is at 
\mbox{$m_\pi/m_\rho =$} \mbox{$0.578$}, which corresponds to the $s$-quark
mass, and thus no physical decay of the $\rho$ can take place, the
amplitude $\langle \rho | \pi \pi \rangle$ is accessible through study 
of state mixing on the lattice.  

Since the width of the $\rho$ is almost 20\% of its mass while the
widths of the $\phi$ and $K^\star$ are significantly less important,
we expect the ladder approximation for the BSE kernel (which omits the
strong channels $\pi \pi$, $K K$ and $K \pi$ respectively) to be less
accurate for the $\rho$ than for the $\phi$ and $K^\star$.
Accordingly we speculate that this is largely the reason why the
result for $g_{\rho \to \pi\pi}$ in Table~\ref{tab:polefit} deviates
from experiment twice as much (15\%) as do the other decay constants.

\section{pQCD Limit of Form Factors}

Besides the soft physical characteristics of light mesons, the present 
DSE model should also reproduce perturbative QCD limits.   This has been
checked for the uv behavior of the quark mass function $M(p^2)$; both
the leading log behavior away from the chiral limit, and the 
coefficient of the leading $1/p^2$ behavior in the chiral limit reproduce
the exact 1-loop results of QCD~\cite{Maris:1997tm}.   A more difficult
task is to test the asymptotic behavior of meson form factors against
pQCD predictions.   This is complicated by the fact that covariant 
ladder-rainbow calculations that link the dressed quark propagator,
the BSE, and the impulse approximation for form factors have only
been carried out in Euclidean metric for practical reasons.   The 
mass-shell constraint for mesons then requires an analytic continuation
which entails complex quark momenta in loop integrals.    This greatly
hinders the asymptotic analysis.   

A case that is free of these difficulties is the symmetric 
$\gamma^* \pi \to \gamma^*$ transition where the photons are taken to have 
equal virtuality $Q^2$ and there is only one mass-shell constraint.
Since $m_\pi^2$ is negligible compared to all other scales in the problem,
all involved quark momenta are essentially real and spacelike.  In
\Fig{fig:piggsymlog} we show the result~\cite{Maris:2002mz} of the present 
DSE model compared to the pQCD asymptotic behavior~\cite{Lepage:1980fj} 
obtained from the light-cone operator product expansion.   (In this case,
log corrections occur at sub-leading order.)   The numerically generated 
asymptotic behavior of the DSE-based model reproduces the pQCD limit as
it must.   By about 2~GeV$^2$ the dressing of the photon vertices becomes
negligible; however the 3-point function does not become an effective
2-point function (thereby generating the required power of $f_\pi$)
until about 15-20~GeV$^2$~\cite{Maris:2002mz}.   Such a high scale for
the onset of pQCD behavior is consistent with an earlier 
observation~\cite{Maris:1998hc} in a DSE-based model study of $F_\pi(Q^2)$.

\section{Discussion}

Recent reviews~\cite{Roberts:2000aa,Alkofer:2000wg} put this model in
a wider perspective and compile results
for both meson and baryon physics, an analysis how
quark confinement is manifest in solutions of the DSEs, and both
finite temperature and finite density extensions.  The question of the
relevance and accuracy of the ladder-rainbow truncation has also received some
attention; it has been shown to be particularly suitable for the flavor
octet pseudoscalar mesons since the next-order contributions to the BSE
kernel, in a quark-gluon skeleton graph expansion, have a significant 
amount of cancellation between repulsive and attractive
corrections~\cite{Bender:1996bb,Bender:2002as}.    The preservation of the 
axial vector WTI is what makes the pseudoscalar 
meson sector a robust and ideal base for parameter fixing;
the rainbow-ladder truncation may be used as a convenience in that sector.
It is hoped that future interplay between lattice simulations and 
continuum modeling will increase our understanding of QCD for hadron physics.

\begin{acknowledgments}
Thanks are due A. Faessler and T. Gutsche for 
organization of a stimulating school in Erice.  Very helpful comments 
from Pieter Maris are acknowledged.   This work was supported in part 
by NSF grants No. PHY-0071361 and INT-0129236.
\end{acknowledgments}
\bibliography{refsPM,refsPCT,refsCDR,refs,refsMAP}

\begin{thebibliography}{31}
\expandafter\ifx\csname natexlab\endcsname\relax\def\natexlab#1{#1}\fi
\expandafter\ifx\csname bibnamefont\endcsname\relax
  \def\bibnamefont#1{#1}\fi
\expandafter\ifx\csname bibfnamefont\endcsname\relax
  \def\bibfnamefont#1{#1}\fi
\expandafter\ifx\csname citenamefont\endcsname\relax
  \def\citenamefont#1{#1}\fi
\expandafter\ifx\csname url\endcsname\relax
  \def\url#1{\texttt{#1}}\fi
\expandafter\ifx\csname urlprefix\endcsname\relax\def\urlprefix{URL }\fi
\providecommand{\bibinfo}[2]{#2}
\providecommand{\eprint}[2][]{\url{#2}}

\bibitem[{\citenamefont{Roberts and Schmidt}(2000)}]{Roberts:2000aa}
\bibinfo{author}{\bibfnamefont{C.~D.} \bibnamefont{Roberts}} \bibnamefont{and}
  \bibinfo{author}{\bibfnamefont{S.~M.} \bibnamefont{Schmidt}},
  \bibinfo{journal}{Prog. Part. Nucl. Phys.} \textbf{\bibinfo{volume}{45S1}},
  \bibinfo{pages}{1} (\bibinfo{year}{2000}), \eprint{nucl-th/0005064}.

\bibitem[{\citenamefont{Maris and Roberts}(1997)}]{Maris:1997tm}
\bibinfo{author}{\bibfnamefont{P.}~\bibnamefont{Maris}} \bibnamefont{and}
  \bibinfo{author}{\bibfnamefont{C.~D.} \bibnamefont{Roberts}},
  \bibinfo{journal}{Phys. Rev.} \textbf{\bibinfo{volume}{C56}},
  \bibinfo{pages}{3369} (\bibinfo{year}{1997}), \eprint{nucl-th/9708029}.

\bibitem[{\citenamefont{Maris et~al.}(1998)\citenamefont{Maris, Roberts, and
  Tandy}}]{Maris:1998hd}
\bibinfo{author}{\bibfnamefont{P.}~\bibnamefont{Maris}},
  \bibinfo{author}{\bibfnamefont{C.~D.} \bibnamefont{Roberts}},
  \bibnamefont{and} \bibinfo{author}{\bibfnamefont{P.~C.} \bibnamefont{Tandy}},
  \bibinfo{journal}{Phys. Lett.} \textbf{\bibinfo{volume}{B420}},
  \bibinfo{pages}{267} (\bibinfo{year}{1998}), \eprint{nucl-th/9707003}.

\bibitem[{\citenamefont{Maris and Tandy}(1999)}]{Maris:1999nt}
\bibinfo{author}{\bibfnamefont{P.}~\bibnamefont{Maris}} \bibnamefont{and}
  \bibinfo{author}{\bibfnamefont{P.~C.} \bibnamefont{Tandy}},
  \bibinfo{journal}{Phys. Rev.} \textbf{\bibinfo{volume}{C60}},
  \bibinfo{pages}{055214} (\bibinfo{year}{1999}), \eprint{nucl-th/9905056}.

\bibitem[{\citenamefont{Atkinson and Johnson}(1988)}]{Atkinson:1988mw}
\bibinfo{author}{\bibfnamefont{D.}~\bibnamefont{Atkinson}} \bibnamefont{and}
  \bibinfo{author}{\bibfnamefont{P.~W.} \bibnamefont{Johnson}},
  \bibinfo{journal}{Phys. Rev.} \textbf{\bibinfo{volume}{D37}},
  \bibinfo{pages}{2296} (\bibinfo{year}{1988}).

\bibitem[{\citenamefont{Roberts and McKellar}(1990)}]{Roberts:1990mj}
\bibinfo{author}{\bibfnamefont{C.~D.} \bibnamefont{Roberts}} \bibnamefont{and}
  \bibinfo{author}{\bibfnamefont{B.~H.~J.} \bibnamefont{McKellar}},
  \bibinfo{journal}{Phys. Rev.} \textbf{\bibinfo{volume}{D41}},
  \bibinfo{pages}{672} (\bibinfo{year}{1990}).

\bibitem[{\citenamefont{Burden et~al.}(1992)\citenamefont{Burden, Roberts, and
  Williams}}]{Burden:1992gd}
\bibinfo{author}{\bibfnamefont{C.~J.} \bibnamefont{Burden}},
  \bibinfo{author}{\bibfnamefont{C.~D.} \bibnamefont{Roberts}},
  \bibnamefont{and} \bibinfo{author}{\bibfnamefont{A.~G.}
  \bibnamefont{Williams}}, \bibinfo{journal}{Phys. Lett.}
  \textbf{\bibinfo{volume}{B285}}, \bibinfo{pages}{347} (\bibinfo{year}{1992}).

\bibitem[{\citenamefont{Krein et~al.}(1992)\citenamefont{Krein, Roberts, and
  Williams}}]{Krein:1992sf}
\bibinfo{author}{\bibfnamefont{G.}~\bibnamefont{Krein}},
  \bibinfo{author}{\bibfnamefont{C.~D.} \bibnamefont{Roberts}},
  \bibnamefont{and} \bibinfo{author}{\bibfnamefont{A.~G.}
  \bibnamefont{Williams}}, \bibinfo{journal}{Int. J. Mod. Phys.}
  \textbf{\bibinfo{volume}{A7}}, \bibinfo{pages}{5607} (\bibinfo{year}{1992}).

\bibitem[{\citenamefont{Maris}(1995)}]{Maris:1995ns}
\bibinfo{author}{\bibfnamefont{P.}~\bibnamefont{Maris}},
  \bibinfo{journal}{Phys. Rev.} \textbf{\bibinfo{volume}{D52}},
  \bibinfo{pages}{6087} (\bibinfo{year}{1995}), \eprint{hep-ph/9508323}.

\bibitem[{\citenamefont{Maris and Tandy}(2000)}]{Maris:2000sk}
\bibinfo{author}{\bibfnamefont{P.}~\bibnamefont{Maris}} \bibnamefont{and}
  \bibinfo{author}{\bibfnamefont{P.~C.} \bibnamefont{Tandy}},
  \bibinfo{journal}{Phys. Rev.} \textbf{\bibinfo{volume}{C62}},
  \bibinfo{pages}{055204} (\bibinfo{year}{2000}), \eprint{nucl-th/0005015}.

\bibitem[{\citenamefont{Maris and Tandy}(2002)}]{Maris:2002mz}
\bibinfo{author}{\bibfnamefont{P.}~\bibnamefont{Maris}} \bibnamefont{and}
  \bibinfo{author}{\bibfnamefont{P.~C.} \bibnamefont{Tandy}},
  \bibinfo{journal}{Phys. Rev.} \textbf{\bibinfo{volume}{C65}},
  \bibinfo{pages}{045211} (\bibinfo{year}{2002}),
  \eprint[http://arXiv.org/abs]{nucl-th/0201017}.

\bibitem[{\citenamefont{Ji and Maris}(2001)}]{Ji:2001pj}
\bibinfo{author}{\bibfnamefont{C.-R.} \bibnamefont{Ji}} \bibnamefont{and}
  \bibinfo{author}{\bibfnamefont{P.}~\bibnamefont{Maris}},
  \bibinfo{journal}{Phys. Rev.} \textbf{\bibinfo{volume}{D64}},
  \bibinfo{pages}{014032} (\bibinfo{year}{2001}), \eprint{nucl-th/0102057}.

\bibitem[{\citenamefont{Jarecke et~al.}(2002)\citenamefont{Jarecke, Maris, and
  Tandy}}]{Jarecke:2002xd}
\bibinfo{author}{\bibfnamefont{D.}~\bibnamefont{Jarecke}},
  \bibinfo{author}{\bibfnamefont{P.}~\bibnamefont{Maris}}, \bibnamefont{and}
  \bibinfo{author}{\bibfnamefont{P.~C.} \bibnamefont{Tandy}}
  (\bibinfo{year}{2002}), \eprint[http://arXiv.org/abs]{nucl-th/0208019}.

\bibitem[{\citenamefont{Bowman et~al.}(2002)\citenamefont{Bowman, Heller, and
  Williams}}]{Bowman:2002bm}
\bibinfo{author}{\bibfnamefont{P.~O.} \bibnamefont{Bowman}},
  \bibinfo{author}{\bibfnamefont{U.~M.} \bibnamefont{Heller}},
  \bibnamefont{and} \bibinfo{author}{\bibfnamefont{A.~G.}
  \bibnamefont{Williams}}, \bibinfo{journal}{Phys. Rev.}
  \textbf{\bibinfo{volume}{D66}}, \bibinfo{pages}{014505}
  (\bibinfo{year}{2002}), \eprint{hep-lat/0203001}.

\bibitem[{\citenamefont{Bowman}(2002)}]{Bowman_privCom02}
\bibinfo{author}{\bibfnamefont{P.}~\bibnamefont{Bowman}}
  (\bibinfo{year}{2002}), \bibinfo{note}{private communication}.

\bibitem[{\citenamefont{Aoki et~al.}(1999)}]{Aoki:1999ff}
\bibinfo{author}{\bibfnamefont{S.}~\bibnamefont{Aoki}} \bibnamefont{et~al.}
  (\bibinfo{collaboration}{CP-PACS}), \bibinfo{journal}{Phys. Rev.}
  \textbf{\bibinfo{volume}{D60}}, \bibinfo{pages}{114508}
  (\bibinfo{year}{1999}), \eprint{hep-lat/9902018}.

\bibitem[{\citenamefont{Allton et~al.}(1999)}]{Allton:1998gi}
\bibinfo{author}{\bibfnamefont{C.~R.} \bibnamefont{Allton}}
  \bibnamefont{et~al.} (\bibinfo{collaboration}{UKQCD}),
  \bibinfo{journal}{Phys. Rev.} \textbf{\bibinfo{volume}{D60}},
  \bibinfo{pages}{034507} (\bibinfo{year}{1999}), \eprint{hep-lat/9808016}.

\bibitem[{\citenamefont{Hawes et~al.}(1998)\citenamefont{Hawes, Maris, and
  Roberts}}]{Hawes:1998cw}
\bibinfo{author}{\bibfnamefont{F.~T.} \bibnamefont{Hawes}},
  \bibinfo{author}{\bibfnamefont{P.}~\bibnamefont{Maris}}, \bibnamefont{and}
  \bibinfo{author}{\bibfnamefont{C.~D.} \bibnamefont{Roberts}},
  \bibinfo{journal}{Phys. Lett.} \textbf{\bibinfo{volume}{B440}},
  \bibinfo{pages}{353} (\bibinfo{year}{1998}), \eprint{nucl-th/9807056}.

\bibitem[{\citenamefont{Fischer}(2002)}]{Fischer_PhD02}
\bibinfo{author}{\bibfnamefont{C.~S.} \bibnamefont{Fischer}}
  (\bibinfo{year}{2002}), \bibinfo{note}{phD Thesis, University of Tuebingen,
  unpublished}.

\bibitem[{\citenamefont{Alkofer}(2002)}]{Alkofer_privcom02}
\bibinfo{author}{\bibfnamefont{R.}~\bibnamefont{Alkofer}}
  (\bibinfo{year}{2002}), \bibinfo{note}{private communication}.

\bibitem[{\citenamefont{Ivanov et~al.}(1999)\citenamefont{Ivanov, Kalinovsky,
  and Roberts}}]{Ivanov:1998ms}
\bibinfo{author}{\bibfnamefont{M.~A.} \bibnamefont{Ivanov}},
  \bibinfo{author}{\bibfnamefont{Y.~L.} \bibnamefont{Kalinovsky}},
  \bibnamefont{and} \bibinfo{author}{\bibfnamefont{C.~D.}
  \bibnamefont{Roberts}}, \bibinfo{journal}{Phys. Rev.}
  \textbf{\bibinfo{volume}{D60}}, \bibinfo{pages}{034018}
  (\bibinfo{year}{1999}), \eprint{nucl-th/9812063}.

\bibitem[{\citenamefont{Maris}(2000)}]{Maris:2000zf}
\bibinfo{author}{\bibfnamefont{P.}~\bibnamefont{Maris}} (\bibinfo{year}{2000}),
  \eprint{nucl-th/0009064}.

\bibitem[{\citenamefont{Flynn and Sachrajda}(1998)}]{Flynn:1998ca}
\bibinfo{author}{\bibfnamefont{J.~M.} \bibnamefont{Flynn}} \bibnamefont{and}
  \bibinfo{author}{\bibfnamefont{C.~T.} \bibnamefont{Sachrajda}},
  \bibinfo{journal}{Adv. Ser. Direct. High Energy Phys.}
  \textbf{\bibinfo{volume}{15}}, \bibinfo{pages}{402} (\bibinfo{year}{1998}),
  \eprint{hep-lat/9710057}.

\bibitem[{\citenamefont{Maris and Tandy}(2001)}]{Maris:2001rq}
\bibinfo{author}{\bibfnamefont{P.}~\bibnamefont{Maris}} \bibnamefont{and}
  \bibinfo{author}{\bibfnamefont{P.~C.} \bibnamefont{Tandy}},
  \bibinfo{journal}{Mesons as Bound States of Confined Quarks: Zero and Finite
  Temperature, for the proceedings of Research Program at the Erwin Schodinger
  Institute on Confinement, Vienna, Austria, 5 May - 17 Jul 2000,}
  (\bibinfo{year}{2001}), \eprint{nucl-th/0109035}.

\bibitem[{\citenamefont{Maris}(2001)}]{Maris:2001am}
\bibinfo{author}{\bibfnamefont{P.}~\bibnamefont{Maris}} (\bibinfo{year}{2001}),
  \eprint[http://arXiv.org/abs]{nucl-th/0112022}.

\bibitem[{\citenamefont{McNeile and Michael}(2002)}]{McNeile:2002fh}
\bibinfo{author}{\bibfnamefont{C.}~\bibnamefont{McNeile}} \bibnamefont{and}
  \bibinfo{author}{\bibfnamefont{C.}~\bibnamefont{Michael}}
  (\bibinfo{collaboration}{UKQCD}) (\bibinfo{year}{2002}),
  \eprint{hep-lat/0212020}.

\bibitem[{\citenamefont{Lepage and Brodsky}(1980)}]{Lepage:1980fj}
\bibinfo{author}{\bibfnamefont{G.~P.} \bibnamefont{Lepage}} \bibnamefont{and}
  \bibinfo{author}{\bibfnamefont{S.~J.} \bibnamefont{Brodsky}},
  \bibinfo{journal}{Phys. Rev.} \textbf{\bibinfo{volume}{D22}},
  \bibinfo{pages}{2157} (\bibinfo{year}{1980}).

\bibitem[{\citenamefont{Maris and Roberts}(1998)}]{Maris:1998hc}
\bibinfo{author}{\bibfnamefont{P.}~\bibnamefont{Maris}} \bibnamefont{and}
  \bibinfo{author}{\bibfnamefont{C.~D.} \bibnamefont{Roberts}},
  \bibinfo{journal}{Phys. Rev.} \textbf{\bibinfo{volume}{C58}},
  \bibinfo{pages}{3659} (\bibinfo{year}{1998}), \eprint{nucl-th/9804062}.

\bibitem[{\citenamefont{Alkofer and von Smekal}(2001)}]{Alkofer:2000wg}
\bibinfo{author}{\bibfnamefont{R.}~\bibnamefont{Alkofer}} \bibnamefont{and}
  \bibinfo{author}{\bibfnamefont{L.}~\bibnamefont{von Smekal}},
  \bibinfo{journal}{Phys. Rept.} \textbf{\bibinfo{volume}{353}},
  \bibinfo{pages}{281} (\bibinfo{year}{2001}),
  \eprint[http://arXiv.org/abs]{hep-ph/0007355}.

\bibitem[{\citenamefont{Bender et~al.}(1996)\citenamefont{Bender, Roberts, and
  Von~Smekal}}]{Bender:1996bb}
\bibinfo{author}{\bibfnamefont{A.}~\bibnamefont{Bender}},
  \bibinfo{author}{\bibfnamefont{C.~D.} \bibnamefont{Roberts}},
  \bibnamefont{and}
  \bibinfo{author}{\bibfnamefont{L.}~\bibnamefont{Von~Smekal}},
  \bibinfo{journal}{Phys. Lett.} \textbf{\bibinfo{volume}{B380}},
  \bibinfo{pages}{7} (\bibinfo{year}{1996}), \eprint{nucl-th/9602012}.

\bibitem[{\citenamefont{Bender et~al.}(2002)\citenamefont{Bender, Detmold,
  Roberts, and Thomas}}]{Bender:2002as}
\bibinfo{author}{\bibfnamefont{A.}~\bibnamefont{Bender}},
  \bibinfo{author}{\bibfnamefont{W.}~\bibnamefont{Detmold}},
  \bibinfo{author}{\bibfnamefont{C.~D.} \bibnamefont{Roberts}},
  \bibnamefont{and} \bibinfo{author}{\bibfnamefont{A.~W.}
  \bibnamefont{Thomas}}, \bibinfo{journal}{Phys. Rev.}
  \textbf{\bibinfo{volume}{C65}}, \bibinfo{pages}{065203}
  (\bibinfo{year}{2002}), \eprint[http://arXiv.org/abs]{nucl-th/0202082}.

\end{thebibliography}

\end{document}